# Use of self-correlation metrics for evaluation of information properties of binary strings

Sergei Viznyuk

**Abstract**

It is demonstrated that appropriately chosen computable metrics based on self-correlation properties provide a degree of determinism sufficient to segregate binary strings by level of information content.

Shannon's classic formula [1]
$$H = -\sum_i p_i \cdot \log_2(p_i) \tag{1}$$
, originally defined as a *measure of information* contained in the message, has been the cornerstone of most works on information theory [3]. Later the definition has been refined [2,3] to the effect that entropy $H$ is rather a measure of *missing information* needed to determine the state of an object. Some valid interpretations of the value provided by (1) are

a) the minimum number of yes/no answers needed to describe the object in terms of predefined set of observables with mutually independent probabilities $p_i$
b) the minimum length to which the message can be compressed, in bits per symbol
c) the minimum channel capacity required to transmit the message, in bits per symbol

The value for entropy provided by (1) depends on probability distribution $p_i$ of the describing observables. In a case of message strings, the set of observables makes up an *alphabet*, or *codebook*. A message and a codebook allow defining the information content of the message using (1). Without known codebook the information content of the message is not defined, despite efforts [4,5,6] towards definition in terms of *algorithmic* (Kolmogorov) complexity or *effective* complexity. It was shown [7] that such definitions always imply a choice of a *codebook*. If we are to involve terminology from quantum mechanics we can draw an analogy between a set of message strings and an ensemble of *open quantum systems*. An open quantum system and its *environment* together can be viewed as a *closed quantum system* which state (information content) can be determined by measurement of certain observables. Similarly, the information content of a message can be determined in the context of a given codebook (environment). Without codebook the information content of the message cannot be measured, similar to an operator in quantum mechanics not having eigenvalue for the given open quantum system. The approach in this case is to consider the *ensemble* of objects. The ensemble is characterized statistically using a set of descriptive observables. The mean values of the observables can be measured (or calculated using known *density matrix),* to provide a level of information about the *ensemble*, though not specific objects.

The reasoning above leads us to believe we are free to choose observable (or computable) parameters for the statistical evaluation of *not directly observable* properties of the objects in an ensemble in general sense, for example for evaluation of information content. The success of the evaluation is measured by degree of determinism between observed parameter values and the property under evaluation.

We consider a large collection of various types of binary computer files as the ensemble of objects. A computable metrics $M_F$ and $D_F$ have been chosen as the parameters to be used for statistical evaluation of information content of the files. We demonstrate that chosen metrics provide a degree of determinism sufficient to segregate arbitrary files by *level of information*

*content*. The degree of determinism increases with file size resulting in reliable segregation of files larger than 10Mbits. The logic in choosing computable metric is based on a premise that non-random messages should have a degree of self-correlation. We start with defining correlation metric $C_R(n)$ as

$$C_R(n) = \sum_{i=1}^{M} B_i \oplus B_{i+n}, \quad 0 \leq n < M \quad (2)$$

, where $B_i$ is $i^{th}$ bit of a binary message string $B$, consisting of $M$ bits with values 0 or 1, and $\oplus$ is logical XOR operator; $B_{i+n} = B_{i+n-M}$ for $i+n>M$. The range of possible values of $C_R(n)$ is 0 to $M$. Next we define metric $M_F(n)=M-2 \cdot C_R(n)$, and aggregate metrics $M_F$ and $D_F$ as

$$M_F = \sum_{n=0}^{M-1} M_F(n) \quad (3)$$

$$D_F = \sum_{n=0}^{M-1} \left( \frac{M_F(n)}{M} \right)^2 - 1 \quad (4)$$

The computation of $M_F$ and $D_F$ metrics has been performed using *corrbits* program (see http://www.phystech.com/download/corrbits.htm) on a large collection of various types of binary computer files:

1. Group 1: Microsoft Excel, Word, PowerPoint, Visio, RTF documents, delimited and position-based non-random data files, Windows and UNIX executables and library files, plain text files, various other types of non-compressed files: TTF, REP, JSP, RDF, TAR, HTM, LOG, XML. The files were collected from different sources. Total number of files in this group 1122. The size of the files in this group ranges from 13 bytes to 4.3Mbytes
2. Group 2: Compressed non-random computer files. BZIP2 and GZIP compression programs have been used to compress files of the same types as in Group 1, and *random slices* of the compressed files as well as whole files have been analyzed with *corrbits* program. Total number of files in this group 1867. The size of the files in this group ranges from 30 bytes to 2.3Mbytes.
3. Group 3: Random data files obtained from /dev/random device on several different UNIX servers. Total number of files analyzed in this group 1334. The size of the files in this group ranges from 14 bytes to 4.5Mbytes.

Figure 1 show values of $M_F$ metric plotted against the size $M$ of the analyzed files.

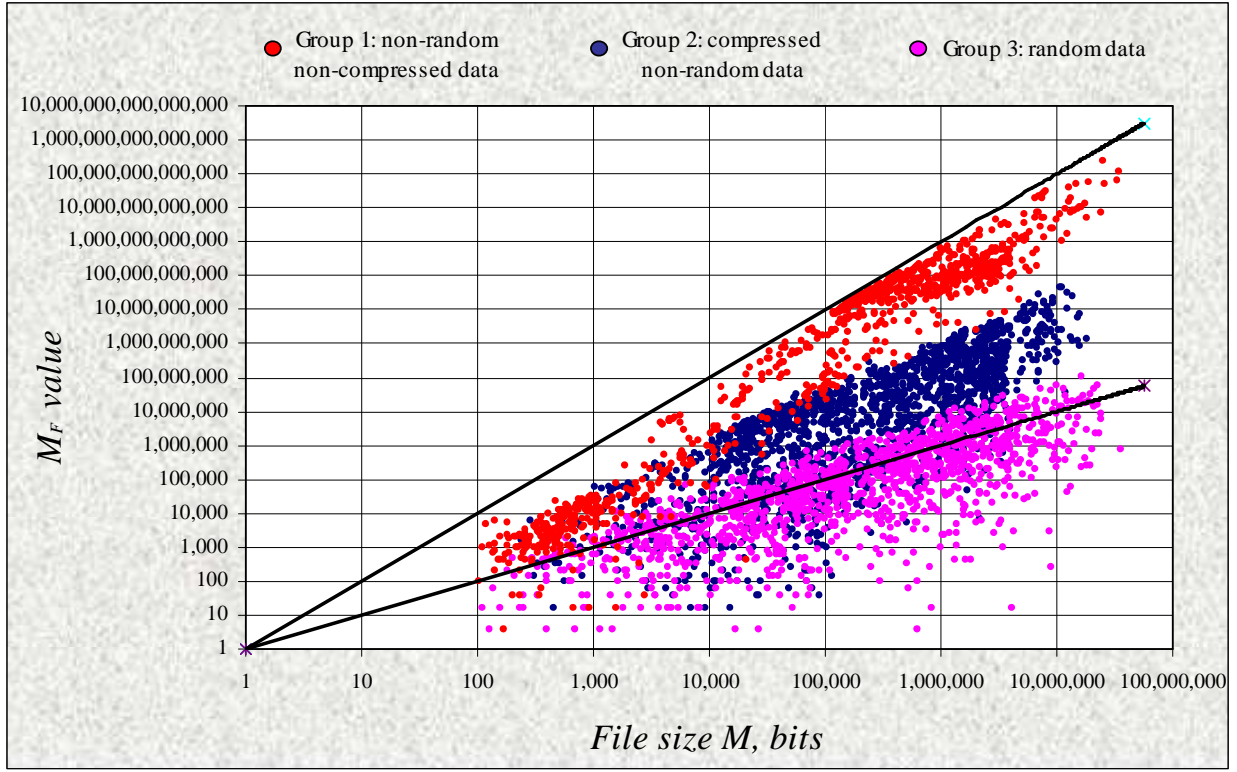

Figure 1

While at smaller file size $M$ the $M_F$ values for all three groups are not clearly segregated, the separation grows with file size and with $M$ greater than 10Mbits the segregation gets close to 100%. The asymptotic behavior of the groupings becomes also evident at bigger file size. $M_F$ values for Group 1 exhibit $M_F \approx M^2$ asymptotic dependency, while for Group 3 it is $M_F \approx M$, and Group 2 approximately follows $M_F \approx M^{3/2}$.

From (3) metric $M_F = M^2$ for strings consisting of all 0 or all 1 bits. The information content of such strings is minimal: 0 according to (1). Therefore the adjustment in the first order to the metric $M_F$ has been made as

$$Adj.M_F = \left(1 - \frac{M_F}{M^2}\right) \cdot M_F \qquad (5)$$

Figure 2 show values of $Adj.M_F$ metric plotted against the size $M$ of the analyzed files.

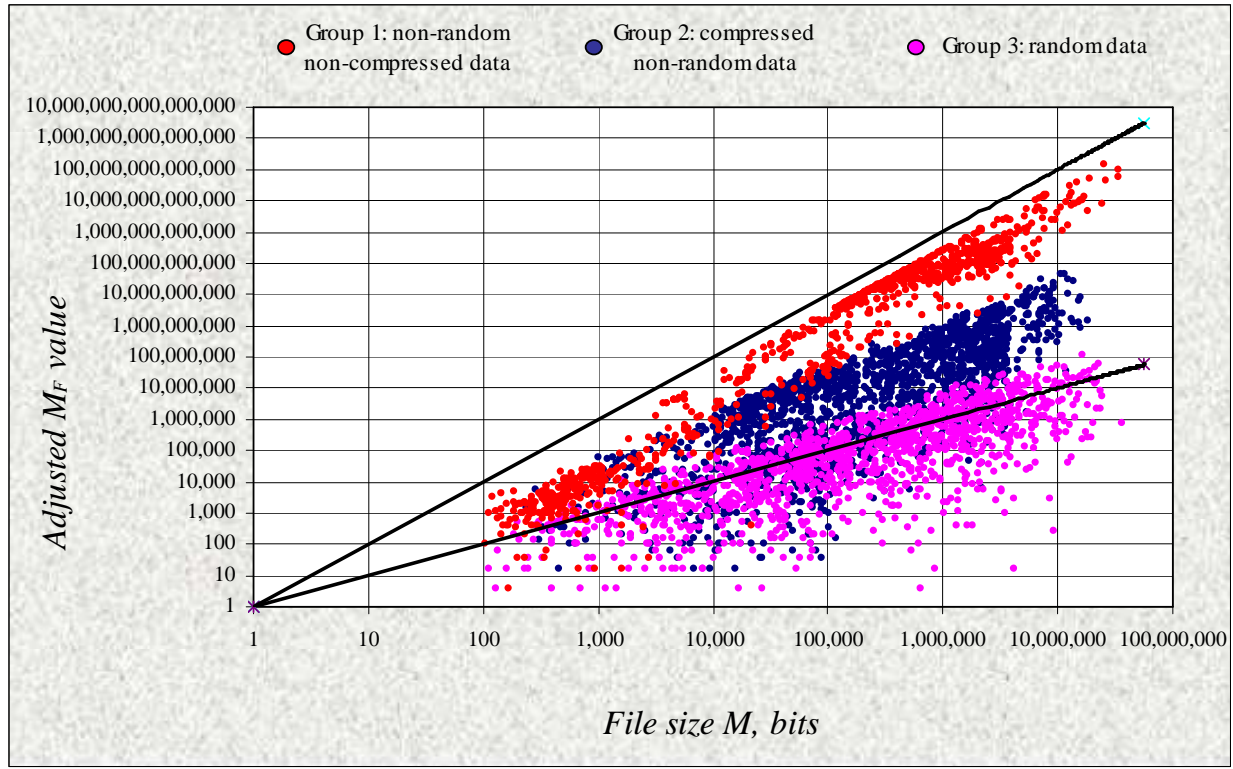

Figure 2

The *Adj.M$_F$* metric shows the same asymptotic behavior as *M$_F$* while potentially allowing more accurate segregation of files by the level of information content. As evident from Figure 2 *Adj.M$_F$* metric breaks the ensemble of arbitrary binary strings into "spectrum" bands, acting as a *dispersion operator*.

Other possible metrics may provide different way of segregating binary message strings. As an example we consider *D$_F$* metric provided by (4). Figure 3 shows values of *D$_F$* metric plotted against the size *M* of the same collection of files as Figures 1-2. Figure 3 demonstrates the qualitative difference in asymptotic behavior of *D$_F$* metric between binary strings in Group 1 and Groups 2-3. We can speculate that the difference in asymptotic behavior of *D$_F$* metric between Group 1 and Groups 2-3 is tied to the amount of what can be considered *self-contained information*. Group1 contains significant amount of self-contained data, Group 3 has none, and Group 2 has very small amount inserted as part of the compression routine metadata.
The asymptotic behavior of *D$_F$* metric for Group 1 follows *D$_F$≈M/100,* and for Group 2-3 *D$_F$≈1*.

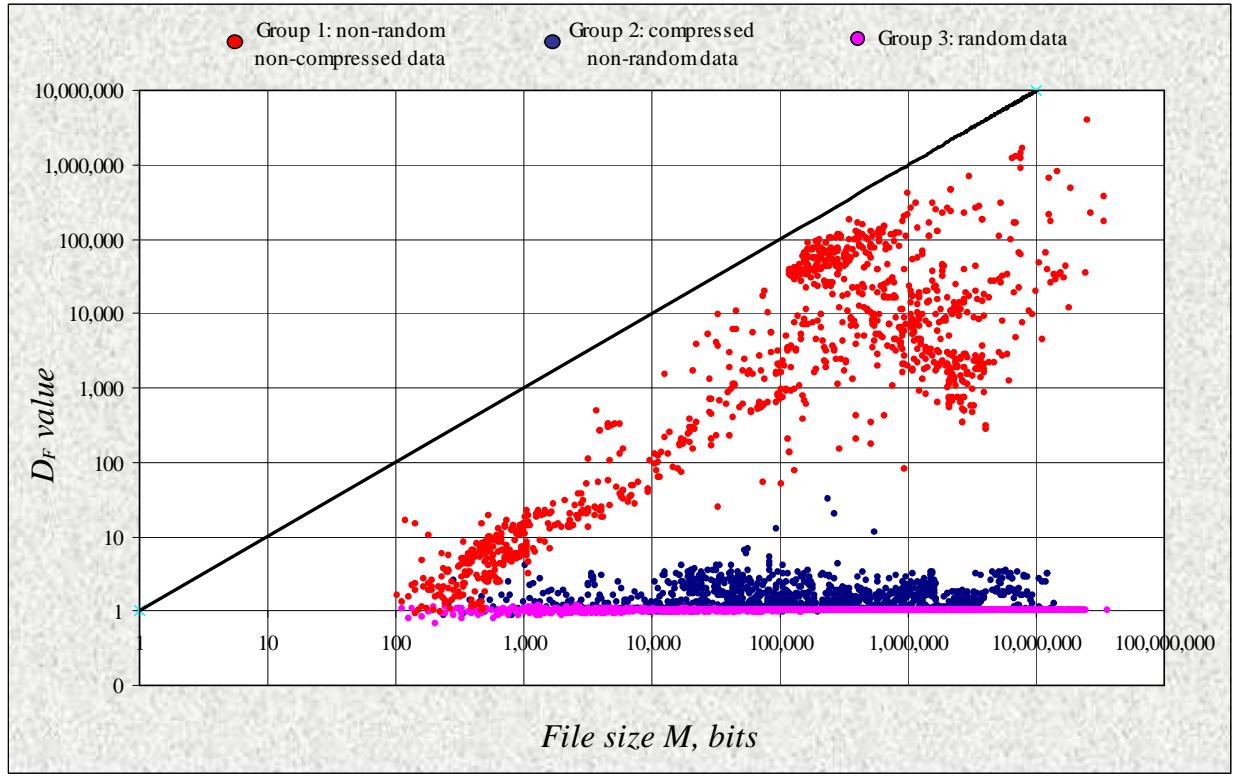

Figure 3

It has been demonstrated that a computable metrics based on self-correlation properties can be used for statistical evaluation of information properties of binary strings. The degree of determinism between the chosen metrics and evaluated property is shown to increase with size of the strings. The particular choice of metrics can be expanded in the future or algorithms behind their calculation modified to improve the practical usefulness of the approach.